# Nonvolatile Control of Nonlinear Hall and Circular Photogalvanic Effects via Berry Curvature Dipole in Multiferroic Monolayer CrNBr$_2$

Wenzhe Zhou[1], Dehe Zhang[1], Guibo Zheng[1], Yinheng Li[1], and Fangping Ouyang[1,2,3,*]

[1]*School of Physics, Hunan Key Laboratory for Super-Microstructure and Ultrafast Process, and Hunan Key Laboratory of Nanophotonics and Devices, Central South University, Changsha 410083, People's Republic of China*

[2]*School of Physics and Technology, Xinjiang Key Laboratory of Solid State Physics and Devices, Xinjiang University, Urumqi 830046, People's Republic of China*

[3]*State Key Laboratory of Powder Metallurgy, and Powder Metallurgy Research Institute, Central South University, Changsha 410083, People's Republic of China*

**Abstract**

The Berry curvature dipole induced by symmetry breaking play a pivotal role in electronic transport properties and nonlinear responses, such as the nonlinear Hall effect and circular photogalvanic effect. The study of the Berry curvature dipole, often explored in time-reversal symmetric systems, but it should not be limited to such materials. Here, we predicted that the ferroelectricity in monolayer CrNBr$_2$ produces Berry curvature dipole, leading to the nonlinear Hall effect and circular photogalvanic current. The linear anomalous Hall effect and circularly polarized optical absorption, governed by spin-orbit coupling, are independent of ferroelectric polarization and exhibit extremely small conductance. In contrast, multiferroic monolayer CrNBr$_2$ achieves a large nonlinear Hall conductivity (~1 $e^3/\hbar \cdot eV^{-1}$ at 30 K) and circular photogalvanic current, despite its suppression at high temperatures from phonon scattering. The coupling between the ferroelectric polarization and the Berry curvature dipoles (intra-band for nonlinear Hall conductance and inter-band for circular photogalvanic current) allows for nonvolatile switching of these effects, presenting substantial promise for nanoelectronic and optoelectronic devices.

*Corresponding author. E-mail address: ouyangfp06@tsinghua.org.cn

# 1. Introduction

Berry curvature has become increasingly significant due to its central role in various quantum phenomena [1, 2], such as the anomalous Hall effect (AHE) [3], valley polarization [4], and chirality in light-matter interactions [5, 6]. Berry curvature acts as an effective magnetic field in momentum space and is responsible for momentum-orbital locking. In topological Chern insulators, the Chern number is defined as the integral of the Berry curvature over the Brillouin zone, which is quantized [3]. Berry curvature induces a transverse velocity for charge carriers under an electric field, serving as the origin of unique transport phenomena such as the AHE. The emergence of Berry curvature is intimately connected to the breaking of inversion symmetry. For instance, in monolayer transition metal dichalcogenides—a prototypical class of valley-polarized materials—the absence of inversion symmetry renders the K and -K valleys inequivalent, endowing them with opposite Berry curvatures. Under spin-orbit coupling (SOC), this further leads to spin-valley locking and consequently imparts optical excitations with distinct circular dichroism [7].

Berry curvature-enabled valley polarization empowers the manipulation of a new degree of freedom and the rise of a new research frontier: valleytronics [8, 9]. The valley polarization properties of various novel two-dimensional materials have been extensively studied, including spin-valley locking in non-magnetic materials [10-20], intrinsic valley polarization in magnetic materials [21-34], and non-volatile control of valley polarization in multiferroic materials [35-40]. On the other hand, the AHE, caused by the anomalous velocity due to Berry curvature, provides a valuable probe for the valley degree of freedom, encompassing phenomena such as the valley Hall effect [41-43].

Recently, the Berry curvature dipole has garnered increasing attention due to its role in nonlinear transport or photogalvanic effects. *I. Sodemann* and coauthors established a theoretical derivation of the nonlinear AHE arising from the Berry curvature dipole in time-reversal symmetric materials [44]. In the nonlinear AHE, the second-harmonic transverse voltage exhibits a quadratic dependence on the bias current, which is significantly larger than the linear response in materials protected by time-reversal symmetry. The exploration of the nonlinear AHE spans a variety of material systems, ranging from bilayer or few-layer WTe$_2$ [45-47], NbIrTe$_4$ [48], TaIrTe$_4$ [49], BaMnSb$_2$ [50], monolayer phosphorene with symmetry breaking [51], CdTe [52], to topological antiferromagnets [53] and moiré systems [54-57]. It is established both experimentally and theoretically that the intrinsic origin of the nonlinear Hall effect lies in the Berry curvature dipole. The nonlinear AHE can also serve as a probe for Berry curvature and topological phase transitions [58], and also enables the detection of Néel vector reversal in antiferromagnetic materials [59].

The Berry curvature dipole comprises intra-band and inter-band distributions, where the intra-band term directly generates nonlinear electrical transport, while the inter-band term gives rise to the nonlinear photovoltaic effect. The circular photogalvanic effect (CPGE) has been observed in materials lacking inversion symmetry [60-68]. These nonlinear responses are governed by symmetry breaking, which naturally directs research attention to the CPGE in ferroelectric materials [69-71], where nonvolatile control of such responses can be realized.

The nonlinear response driven by the Berry curvature dipole has so far been realized almost exclusively in time-reversal-symmetric systems—a strategy adopted to suppress the linear component arising from the Berry curvature. This approach, however, may be overly restrictive. In spin-polarized bands, certain spatial symmetries, when combined with SOC, can give rise to valley polarization while yielding only a minimal linear anomalous Hall response, thereby preserving the conditions necessary for observing the nonlinear signal. Therefore, we aim to identify multiferroic materials that enable nonvolatile control of spin-polarized nonlinear phenomena, specifically the AHE and CPGE. Inspired by the recently reported multiferroic material VOCl$_2$ [72, 73], we predict ferroelectricity-coupled Berry curvature dipole and consequent nonlinear AHE and CPGE in monolayer CrNBr$_2$.

## 2. Methods and computational details

The relaxation and the electronic structures of monolayer CrNBr$_2$ are calculated through the first-principles package DS-PAW software, which is integrated into the visualization software Device Studio. It is also convenient for studying the electronic properties of various materials [74]. The projector augmented wave (PAW) scheme is used, and the energy cutoff of the plane-wave basis is set to be 600 eV [75]. The exchange-correlation effect is treated by the generalized gradient approximation (GGA) of Perdew-Burke-Ernzerhof (PBE) functional [76]. The structural optimization is stopped after the force on each atom is less than 0.001 eV/Å, and the energy criterion of electronic self-consistent iteration is $10^{-6}$ eV. The Γ-centered k-point grid of 21 × 21 × 1 is adopted. A vacuum region of at least 20 Å is applied along the z-direction to ignore the interaction between repeated slabs. The on-site Coulomb interaction for the d orbital of the Cr atom is added with the parameter U = 3 eV [77]. The phonon spectrum is calculated using the density functional perturbation theory (DFPT) method [78]. A 7 × 7 × 1 superlattice is used to calculate the dynamical matrix. The first-principles calculations of Berry curvature in reciprocal space and the Wannier fitting are carried out as implemented in the VASP+Wannier90 package [79].

## 3. Results and discussion

The structure of monolayer $CrNBr_2$ is shown in **Fig. 1(a, b)**. Each Cr atom bonds with two nearest-neighbor N atoms and four Br atoms. For the lowest-energy ground-state structure, the Cr atom deviates from the center of the two N atoms, resulting in broken inversion symmetry. The opposite-direction displacements of the Cr atoms induce opposing electric polarizations along the x-axis, leading to ferroelectricity. The lattice constants of monolayer $CrNBr_2$ are optimized to be $a$ = 3.797 Å and $b$ = 3.634 Å. The dynamic and thermodynamic stabilities of the ferroelectric $CrNBr_2$ monolayer are verified. The phonon spectrum and result of molecular dynamics are shown in **Fig. 1(c, d)**. An acoustic branch has a small imaginary frequency near the Γ point, which is caused by the insufficient accuracy of the dynamical matrix. It can be eliminated by calculating a larger supercell and does not indicate material instability. The absence of other imaginary frequencies confirms that the monolayer $CrNBr_2$ is dynamically stable. The average energy per atom varies in the range of $3/2k_BT$, indicating that monolayer $CrNBr_2$ is thermodynamically stable. The supercell structure exhibits no significant changes after 10 ps of dynamics simulation at room temperature, as shown in the inset of **Fig. 1(d)**.

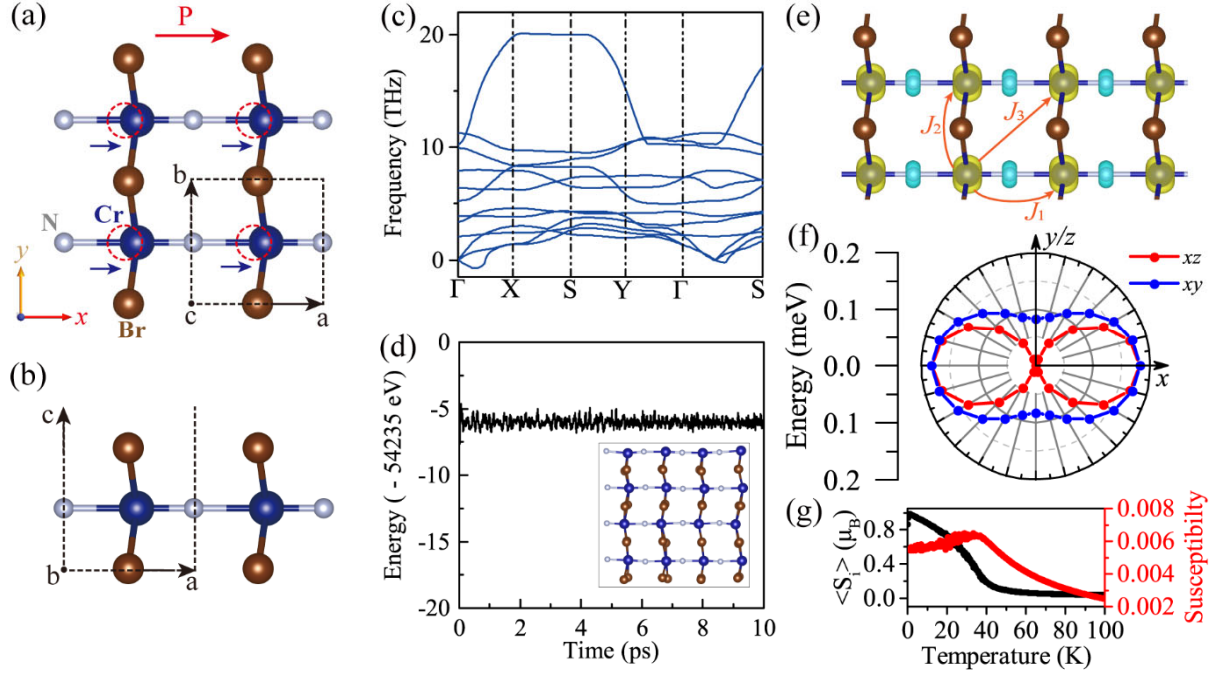

**Fig. 1** The top (a) and side (b) views of monolayer CrNBr$_2$ with the ferroelectric polarization. (c) The phonon spectrum of monolayer CrNBr$_2$. (d) Variation of energy per unit cell of monolayer CrNBr$_2$ with time during ab initio molecular simulation at room temperature. The inset shows the final structure after 10 ps at 300 K. (e) The spin-polarized charge density of monolayer CrNBr$_2$. Yellow and cyan isosurfaces indicate the spin-up and spin-down charges. (f) The energy per cell of monolayer CrNBr$_2$ under different magnetization directions. The energy of magnetization along the positive z-axis direction is set to 0 as a reference. (g) The changes of magnetic moment and magnetic susceptibility as functions of temperature in the Monte Carlo simulation of monolayer CrNBr$_2$.

The intrinsic magnetic properties of monolayer CrNBr$_2$ have been determined. Monolayer CrNBr$_2$ exhibits ferromagnetic ordering in its ground state, with the spin-polarized charge density shown in **Fig. 1(e)**. The nearest-neighbor ($J_1$, $J_2$) and next-nearest-neighbor ($J_3$) spin exchange interactions were calculated to be $J_1$ = -35.394 meV, $J_2$ = -4.484 meV, and $J_3$ = -1.991 meV. The magnetic anisotropy energy of monolayer CrNBr$_2$ was also calculated, as shown in **Fig. 1(f)**. The easy magnetization axis of monolayer CrNBr$_2$ lies along the out-of-plane direction, and a certain degree of in-plane anisotropy is also present. Monte Carlo simulations were performed to determine the Curie temperature of monolayer CrNBr$_2$. The temperature dependence of magnetic moment and magnetic susceptibility is shown in **Fig. 1(g)**. The monolayer CrNBr$_2$ undergoes a ferromagnetic-to-paramagnetic transition at approximately

34.67 K. The computational details of the magnetic properties can be found in the Supplementary Materials (SM) [80].

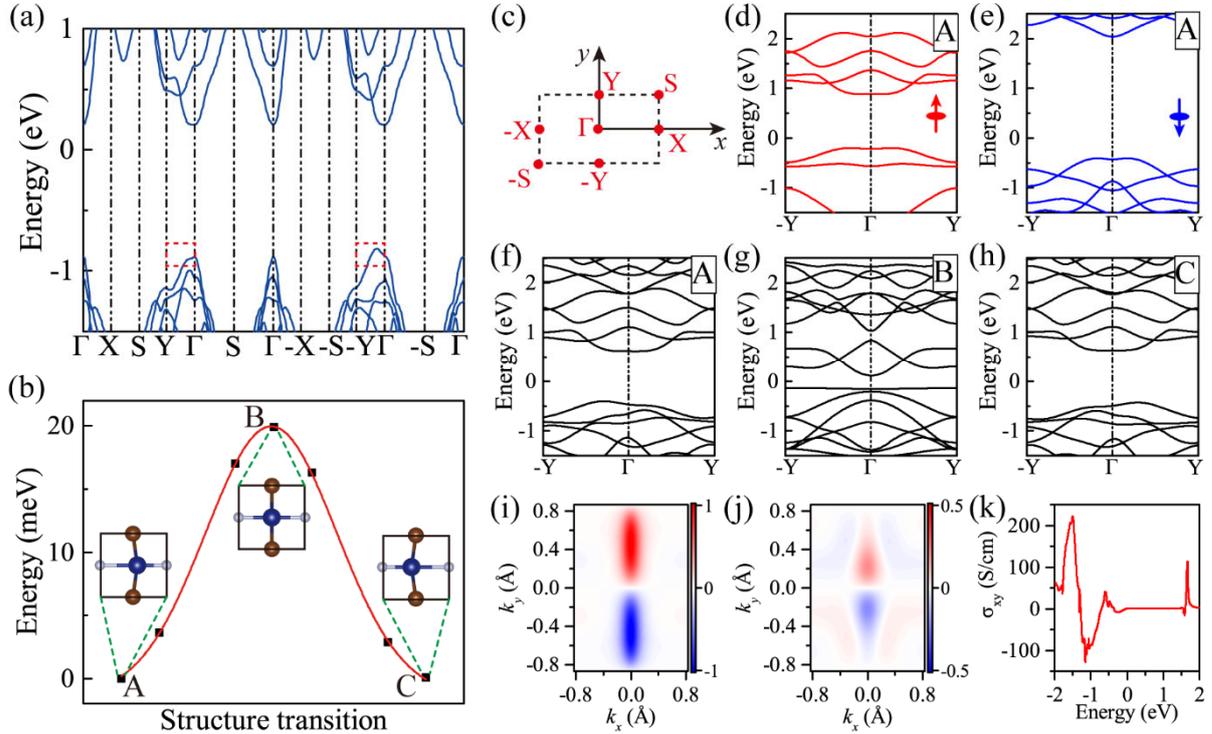

**Fig. 2** (a) The band structure of monolayer CrNBr$_2$ with SOC. (b) Evolution of the energy per cell of monolayer CrNBr$_2$ transforming from one polarization to the opposite polarization. (c) The Brillouin zone and high symmetry point in the reciprocal space of monolayer CrNBr$_2$. The spin-up (d) and spin-down (e) band structures without SOC along the -Y→Γ→Y path. The band structures with SOC of monolayer CrNBr$_2$ along the -Y→Γ→Y path with the structures of A (f), B (g), and C (h). The structures are labelled and shown in (b). The Berry curvature in reciprocal space for occupied states of spin-up (i) and spin-down (j) bands, calculated using the VASP+Wannier90 package. (k) The anomalous Hall conductivity of monolayer CrNBr$_2$ calculated using the VASP+Wannier90 package.

The symmetry breaking in monolayer CrNBr$_2$ leads to inequivalent electronic states at $k$ and -$k$ points in reciprocal space, analogous to valley materials. The band structure of monolayer CrNBr$_2$ with SOC considered is shown in **Fig. 2(a)**. Monolayer CrNBr$_2$ is a ferromagnetic semiconductor. In the absence of SOC, monolayer CrNBr$_2$ exhibits a spin-up bandgap of 1.084 eV and a spin-down bandgap of 2.447 eV, as shown in **Fig. 2(d, e)**. The valence bands demonstrate distinct Mexican-hat-shaped dispersion characteristics, which is a hallmark feature observed in various two-dimensional materials [81]. Without SOC, the $k$ and -$k$ points remain energy-degenerate. However, when SOC is included, the band structure along Γ→Y shows a distinct difference from that along Γ→-Y. This spin-orbit-coupling-induced degeneracy lifting is fundamentally equivalent to the intrinsic valley polarization caused by ferroelectric polarization in multiferroic materials. The structural transition between the two polarization states in monolayer CrNBr$_2$ is depicted in **Fig. 2(b)**. The two polarized structures are labeled A and C, and the symmetric intermediate structure is denoted as B. The energy barrier for the structural transition is about 19.97 meV. The valley polarization of structures A and C is opposite (see **Fig. 2(f-h)**), enabling monolayer CrNBr$_2$ for nonvolatile control of ferrovalley polarization.

The ferroelectric-tunable valley polarization originates from the nonzero Berry curvature coupled with ferroelectricity. The Berry curvature in reciprocal space calculated by VASP+Wannier90 is presented in **Fig. 2(i, j)**. A Berry curvature of significant magnitude is observed along the -Y→Γ→Y path, exhibiting opposite signs between $k_y$ and -$k_y$ points. Similar to ferrovalley materials, the two valleys remain degenerate with identical carrier distributions when SOC is neglected, leading to cancellation of the anomalous Hall conductivity. However, when SOC is included, the degeneracy lifts, resulting in valley-contrasting carrier concentrations and a finite anomalous Hall conductivity. The calculated anomalous Hall conductivity for monolayer CrNBr$_2$ is presented in **Fig. 2(k)**.

The opposite Berry curvatures at -$k_y$ and $k_y$ in reciprocal space suggest that monolayer CrNBr$_2$ may exhibit a significant Berry curvature dipole, which could give rise to nonlinear Hall effect and CPGE. To elucidate the ferroelectricity-driven Berry curvature dipole and its resulting nonlinear Hall effect and CPGE in monolayer CrNBr$_2$, we constructed a tight-binding model based on first-principles calculations using VASP+Wannier90. The detailed parameters of the tight-binding model are provided in the SM [80]. Despite quantitative differences between tight-binding and first-principles calculated band structures, the essential signatures of ferroelectricity-driven Berry curvature exhibit robust agreement.

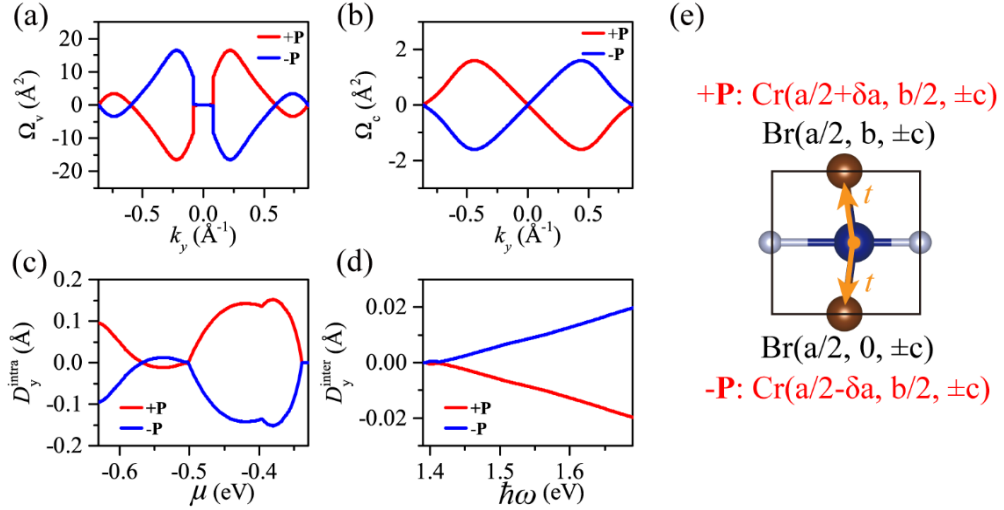

**Fig. 3** The calculated Berry curvatures along the -Y→Γ→Y path for the top of the valence band (a) and the bottom of the conduction band (b) of the spin-up bands of monolayer CrNBr$_2$. (c) The intra-band Berry curvature dipole as a function of the Fermi level for the spin-up bands of monolayer CrNBr$_2$. (d) The inter-band Berry curvature dipole as a function of the photon energy for the spin-up bands of monolayer CrNBr$_2$. The red and blue curves correspond to the structures with positive ferroelectric polarization (Structure A) and negative ferroelectric polarization (Structure C), respectively. (e) Illustration of the two-band model for understanding the ferroelectricity-coupled Berry curvature in monolayer CrNBr$_2$.

**Fig. 3** compares the Berry curvature and its dipole of monolayer CrNBr$_2$ for two opposing ferroelectric polarization states. The Berry curvature distribution satisfies the relation $\Omega(k_y) = -\Omega(-k_y)$ and $\Omega^{+P}(k_y) = -\Omega^{-P}(k_y)$. The Berry curvature near the top of the valence band is significantly larger than that at the bottom of the conduction band, and the reversal of ferroelectric polarization also induces sign inversion of the Berry curvature dipole. Ferroelectricity-driven Berry curvature reversal enables nonvolatile tunability of nonlinear AHE and CPGE. The spin-down bands exhibit similar characteristics, albeit with reduced Berry curvature and a wider bandgap, as illustrated in **Fig. S3** of the SM [80]. This necessitates higher-frequency optical excitation, which is detrimental for device applications.

The ferroelectricity-coupled Berry curvature can be understood by the two-band model of monolayer CrNBr$_2$. As depicted in **Fig. 3(e)**, we only consider the interaction between Cr and Br atoms (denoted as $t$). Under opposite ferroelectric polarizations, the Cr atoms deviate in opposite directions by $p\delta a$, where $p = \pm 1$ represents positive and negative ferroelectric polarizations, respectively. The Hamiltonian can be written as,

$$\hat{H} = \begin{bmatrix} -\dfrac{\Delta}{2} & te^{i[k_x(-p\delta a)+k_y b/2]} + te^{i[k_x(-p\delta a)-k_y b/2]} \\ te^{-i[k_x(-p\delta a)+k_y b/2]} + te^{-i[k_x(-p\delta a)-k_y b/2]} & \dfrac{\Delta}{2} \end{bmatrix}, \quad (1)$$

where Δ denotes the difference in on-site energy. Following the diagonalization of the Hamiltonian, the Berry curvature for the occupied bands can be evaluated using **Eq. (S5)** in the SM [80], as follows,

$$\Omega_1(\mathbf{k}) = -\frac{|C_1|^2 |C_2|^2 16\cos^3[k_y b/2]\sin[k_y b/2](p\delta a)b}{(2+2\cos(k_y b))^2}, \quad (2)$$

where $C_1$ and $C_2$ are the normalization constants for the wave functions. The calculation results for the model confirm the ferroelectricity-driven reversal of the Berry curvature.

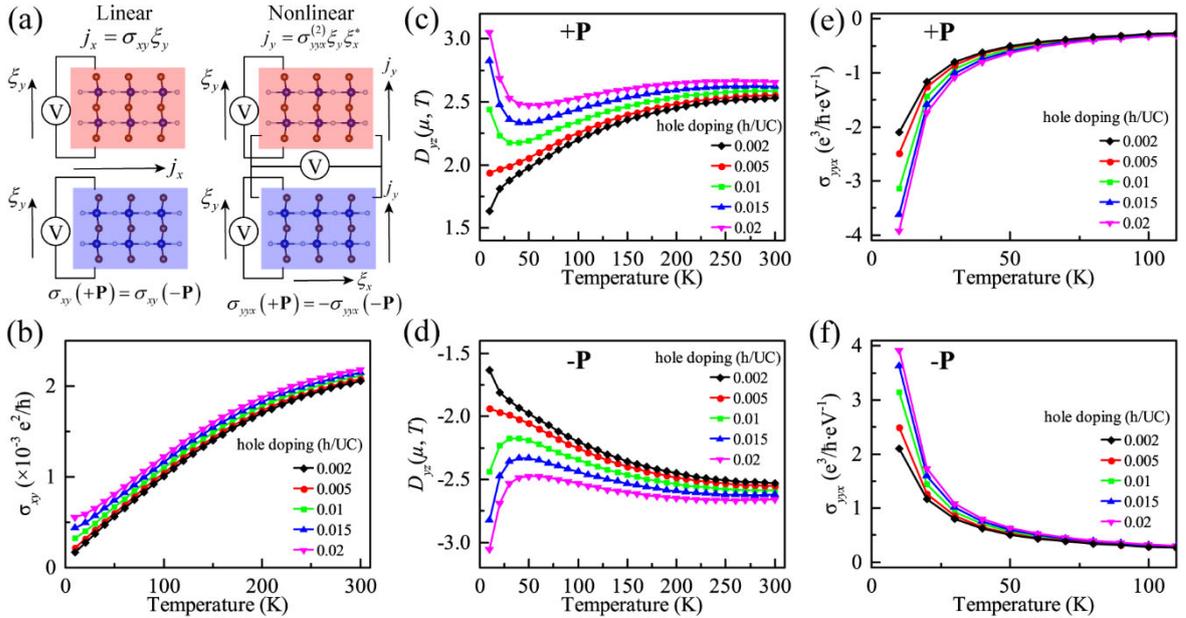

**Fig. 4** (a) The schematic illustration of the linear and nonlinear AHE in monolayer CrNBr$_2$. (b) The calculated temperature dependent linear anomalous Hall conductivity under varying hole doping concentrations with SOC strength of 0.15. Berry curvature dipole as a function of temperature in monolayer CrNBr$_2$ with +P (c) and -P (d) ferroelectric polarizations at different hole doping levels. Temperature-dependent nonlinear anomalous Hall conductivity of monolayer CrNBr$_2$ with +P (e) and -P (f) ferroelectric polarizations at varying hole doping concentrations.

The linear and nonlinear AHE in monolayer CrNBr$_2$ are demonstrated, as presented in **Fig.**

**4**. The linear AHE, where a *y*-directed electric field produces an *x*-directed current, originates intrinsically from the net Berry curvature summed over the Brillouin zone. SOC is essential for the linear AHE, and for monolayer CrNBr$_2$, the linear AHE is very small. An increase in temperature leads to a higher carrier concentration, thereby resulting in a modest enhancement of the AHE. Opposing ferroelectric polarizations result in both opposite energy non-degeneracies and Berry curvatures, they ultimately produce identical linear AHE. The nonlinear AHE, manifested as a *y*-directional current dependent on both in-plane electric field components, arises from the intra-band Berry curvature dipole. **Fig. 4(c, d)** present the temperature-dependent intra-band Berry curvature dipole in monolayer CrNBr$_2$. Notably, a monotonic enhancement of the dipole with temperature is observed at low doping concentrations, in contrast to its non-monotonic behavior, characterized by an initial decrease followed by an increase, at high doping levels. The nonlinear AHE exhibits a pronounced decrease with temperature (as shown in **Fig. 4(e, f)**), a behavior primarily attributed to enhanced phonon scattering. Its magnitude, however, is still notably large relative to the linear effect and experimentally reported values. Crucially, the nonlinear AHE is independent of SOC, and its non-volatile character offers a distinct advantage for controllable device operation via ferroelectric polarization switching.

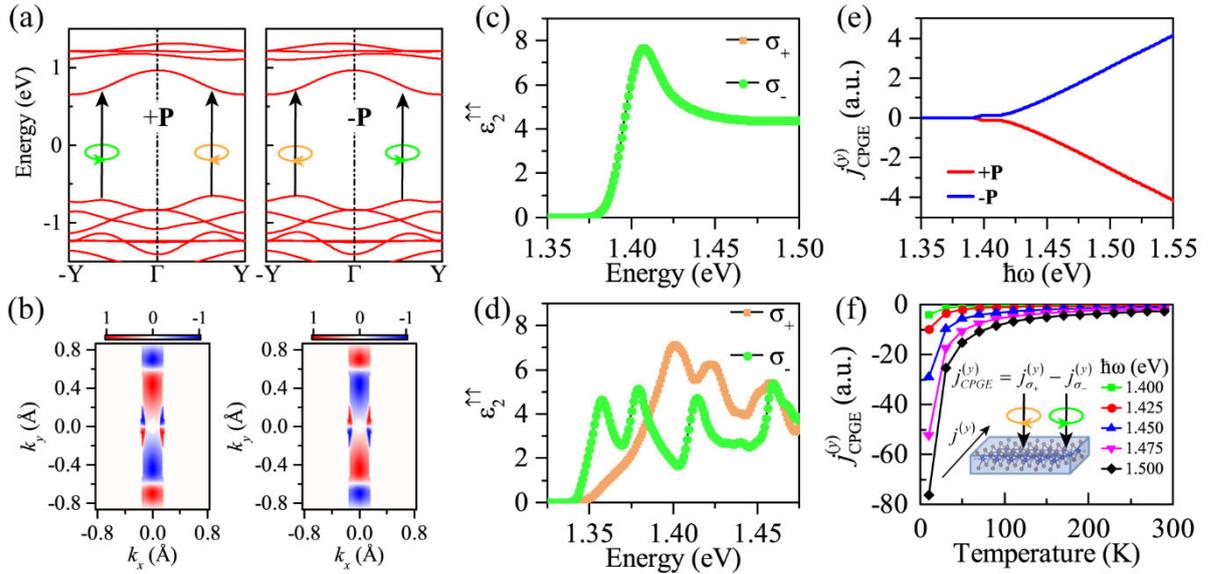

**Fig. 5** (a) Band structures of monolayer CrNBr$_2$ with SOC strength of 0.15 for +P (left) and -P (right) ferroelectric polarizations. (b) The circular polarization of excitation from the top of the valence band to the bottom of the conduction band for monolayer CrNBr$_2$ for +P (left) and -P (right) ferroelectric polarizations. The imaginary part of the complex dielectric function excited by circularly polarized light for monolayer CrNBr$_2$ without (c) and with (d) SOC. (e) The current of CPGE under room temperature for monolayer CrNBr$_2$ for opposite ferroelectric polarizations. (f) The current of CPGE as functions of temperature. The inset is a schematic diagram of the CPGE.

The circular dichroism of the monolayer CrNBr$_2$ is shown in **Fig. 5**. In the absence of SOC, the energies at $k_y$ and $-k_y$ are degenerate, and the circular polarization of the excitation obeys $\eta(k_y) = -\eta(-k_y)$. As a result, the absorption is identical for left- and right-handed circularly polarized light. The inclusion of SOC lifts the energy degeneracy and breaks the odd symmetry of the circular polarization. Consequently, distinct optical band gaps emerge for left- and right-handed circularly polarized light.

The direct origin of circular dichroism arising from SOC is the Berry curvature, which can be measured via photoluminescence. But it is not contingent on the orientation of ferroelectric polarization. On the other hand, the nonlinear response induced by the Berry curvature dipole can be characterized through the CPGE. We calculated the current of CPGE with methodological details provided in the SM [80]. As shown in **Fig. 5(e)**, the current exhibits complete consistency with the inter-band Berry curvature dipole, indicating its potential for nonvolatile manipulation. Considering the temperature dependence of the circular photogalvanic current, similarly, due to the reduction in relaxation time caused by phonon scattering, the current exhibits rapid attenuation with increasing temperature.

## 4. Conclusion

In summary, we predicted the multiferroicity in monolayer CrNBr$_2$, along with the ferroelectricity-driven Berry curvature and its dipole, enabling nonvolatile control of the nonlinear AHE and the CPGE. The monolayer multiferroic material CrNBr$_2$ exhibits an out-of-plane easy magnetization axis with a Curie temperature of 34.67 K, and the energy barrier for ferroelectric switching is approximately 19.97 meV. In monolayer CrNBr$_2$, the top valence band exhibits a Mexican-hat-like dispersion. The two valleys at opposite momenta possess opposite

Berry curvatures. Under spin-orbit coupling, the energy degeneracy between these two valleys is lifted. Consequently, the linear AHE emerges solely due to the presence of SOC and is not modulated by the direction of ferroelectric polarization. The Berry curvature exhibits an odd function distribution with respect to $k_y$, giving rise to a Berry curvature dipole. The intra-band Berry curvature dipole directly gives rise to the nonlinear Hall conductivity. Although phonon scattering can reduce this nonlinear conductivity, its magnitude remains significantly larger than that of the linear Hall effect. Similarly, SOC induces a disparity in circularly polarized optical absorption, yet this effect does not vary with the ferroelectric polarization direction. The inter-band Berry curvature dipole gives rise to a significant CPGE, although it is also limited by phonon scattering. The Berry curvature dipole identified in the predicted monolayer multiferroic $CrNBr_2$ enables non-volatile manipulation of the nonlinear AHE and the CPGE, thereby offering promising prospects for the design and implementation of nonlinear nanoelectronic and optoelectronic devices.

## Conflict of interest

The authors declare that they have no conflict of interest.

## Acknowledgments

This work is financially supported by the National Natural Science Foundation of China (Grant No. 52073308, No. 12004439, No. 12164046, and No. 12304097), the Key Project of the Natural Science Program of Xinjiang Uygur Autonomous Region (Grant No. 2023D01D03), the Tianchi-Talent Project for Young Doctors of Xinjiang Uygur Autonomous Region (No. 51052300570), the Outstanding Doctoral Student Innovation Project of Xinjiang University (No. XJU2023BS028), Hunan Provincial Natural Science Foundation of China (Grant No. 2023JJ40703), and the State Key Laboratory of Powder Metallurgy at Central South University. This work was carried out in part using computing resources at the High Performance Computing Center of Central South University. We also gratefully acknowledge HZWTECH for providing computational facilities and technical support.

## Appendix A. Supplementary materials

Supplementary materials to this article can be found online.